\documentclass[%
 reprint,
 amsmath,amssymb,
 aps,
]{revtex4-2}
\usepackage{graphicx}% Include figure files
\usepackage{dcolumn}% Align table columns on decimal point
\usepackage{bm}% bold math
\usepackage{hyperref}
\usepackage{amsmath}
\hypersetup{hypertex=true,
            colorlinks=true,
            linkcolor=blue,
            anchorcolor=blue,
            citecolor=blue}
\begin{document}

\preprint{APS/123-QED}

\title{Inference of Parameters for Back-shifted Fermi Gas Model \\using Feedback Neural Network}% Force line breaks with \\
\author{Peng-Xiang Du, Tian-Shuai Shang, Kun-Peng Geng, Jian Li}
\email{E-mail:jianli@jlu.edu.cn}
 \affiliation{College of Physics, Jilin University, Changchun 130012, China.}%Lines break automatically or can be forced with \\
\author{Dong-Liang Fang}
\affiliation{%
    Institute of Modern Physics, Chinese Academy of Sciences, Lanzhou 730000, China
}%

%\affiliation{
%\affiliation{
%\affiliation{%


\date{\today}% It is always \today, today,

\begin{abstract}
The back-shifted Fermi gas model is widely employed for calculating nuclear level density (NLD) as it can effectively reproduce experimental data by adjusting parameters. However, selecting parameters for nuclei lacking experimental data poses a challenge. In this study, the feedforward neural network (FNN) was utilized to learn the level density parameters at neutron separation energy $a(S_{n})$ and the energy shift $\varDelta$ for 289 nuclei. Simultaneously, parameters for nearly 3000 nuclei are provided through the FNN. Using these parameters, calculations were performed for neutron resonance spacing in $s$ and $p$ waves, cumulative number of levels, and NLD. The FNN results were also compared with the calculated outcomes of the parameters from fitting experimental data (local parameters) and those obtained from systematic studies (global parameters), as well as the experimental data. 
The results indicate that parameters from the FNN achieve performance comparable to local parameters in reproducing experimental data. Moreover, for extrapolated nuclei, parameters from the FNN still offer a robust description of experimental data.
%\begin{description}
%\end{description}
\end{abstract}

\maketitle

%\tableofcontents

\section{\label{sec:1}introduction}

 Nuclear level density constitutes a fundamental input parameter in the calculation of nuclear reaction cross sections using statistical models, such as the Hauser-Feshbach formula~\cite{hauser1952inelastic,goriely2008improved}. Specifically, in instances where discrete-level information is incomplete or unavailable at a given excitation energy, the level density is employed to calculate the corresponding transmission coefficients. In nuclear reactions involving compound nuclear processes, including capture reactions and fission reactions, knowledge of the level density is of crucial importance~\cite{rajasekaran1981nuclear,yalccin2017cross}. It significantly impacts the calculations of nucleosynthesis in various astrophysical environments and also has substantial implications for the assessment of nuclear fission yields.

NLD is a physical quantity whose direct measurement is challenging. Meanwhile, the most abundant data are from indirect measurements including the cumulative number of discrete levels at low excitation energy and neutron resonance spacing. As an alternative, the Oslo method, progressively developed in recent years, can extract $\gamma$ strength function and level density from the $\gamma$ spectra of specific reactions~\cite{schiller2000extraction,spyrou2014novel,ingeberg2020first,wiedeking2021independent}. However, the NLD data from this technique remains limited.

The earliest theoretical calculations of NLD conducted by Hans Bethe provided a simple analytical expression for the level density using the non-interacting Fermi gas model~\cite{bethe1936attempt,bethe1937nuclear}. Based on such calculations and considering shell effects and pairing effects, a series of phenomenological models have been developed, including the constant temperature model (CTM)~\cite{gilbert1965composite}, the back-shifted Fermi gas model (BFM)~\cite{dilg1973level}, and the generalized superfluid model (GSM)~\cite{ignatyuk1975phenomenological}. Meanwhile, in the past few decades, various microscopic methods based on mean-field approximation or shell models have also been developed to describe NLD. Methods based on mean-field approximations include combinatorial methods~\cite{hilaire2001combinatorial,hilaire2006global,goriely2008improved1,hilaire2012temperature,uhrenholt2013combinatorial,geng2023calculation,jiang2023nuclear} and microscopic statistical methods~\cite{choudhury1977nuclear,goriely1996new,agrawal1998excitation,demetriou2001microscopic,zhao2020microscopic,zhang2023level}. Calculations based on the shell model have also been attempted~\cite{zelevinsky1996nuclear,wang2023projected}. To avoid the substantial computational complexity associated with diagonalizing a massive Hamiltonian, many methods such as Monte Carlo shell model method~\cite{ormand1997estimating,alhassid1999particle,white2000shell,alhassid2003nuclear}, stochastic estimation method~\cite{shimizu2016stochastic,chen2023shell}, moments method~\cite{sen2013high,zelevinsky2019nuclear} and the Lanczos method~\cite{ormand2020microscopic} have been developed. Shell model methods are constrained by the complexity of diagonalizing the Hamiltonian, making it challenging to conduct large-scale calculations for the entire nuclide chart. Combinatorial methods have now achieved a precision level close to that of phenomenological models. Recently, a new approach based on the boson expansion of QRPA excitations has been proposed~\cite{hilaire2023new}. Despite the availability of numerous microscopic methods for calculating NLD, phenomenological models, owing to their simplicity and rapid computational capabilities, continue to be widely employed in various nuclear reaction calculations. For nuclei with discrete levels and $s$-wave neutron resonance spacings ($D_{0}$), phenomenological models can effectively reproduce experimental data by adjusting parameters. According to recent results of fitting parameter, the BFM exhibits slightly better performance in reproducing experimental data on $D_{0}$ compared to CTM and GSM~\cite{koning2008global}.

The BFM primarily utilizes two adjustable parameters, i.e., level density parameter $a$ and energy shift $\varDelta$ to obtain NLD. In earlier fitted results, the $\varDelta$ was found to be negative for all odd-odd nuclei, corresponding to a downward shift of the ground-state energy by 1-3 MeV, it is referred to as the back-shifted Fermi gas model~\cite{dilg1973level,capote2009ripl}. In the subsequent developments, to account for the damping of shell effects with increasing excitation energy, the level density parameter $a$ was modified to be a function of excitation energy~\cite{ignatyuk1975phenomenological}. To address the divergence issue of the original BFM at low energies, an extra term, denoted as $\rho_{0}$ which is related to the level density parameter $a$, energy shift $\varDelta$ and spin cut-off parameter $\sigma^2$, has been added to the expression for the total nuclear level density~\cite{grossjean1985level,demetriou2001microscopic}. Based on these adjustments, the level density parameters (local parameters), $a(S_n)$ ($a$ at the neutron separation energy $S_{n}$) and $\varDelta$,  for 289 nuclei  were determined by fitting experimental data on $D_{0}$ and discrete levels~\cite{koning2008global}. Subsequently, a set of global parameters are derived through systematic investigation to describe NLD of nuclei lacking experimental data~\cite{koning2008global}. The determined local parameters through fitted experimental data exhibit complex variations, and also implicitly reflect shell effects, pairing effects and other effects. Therefore, it is crucial to adopt reliable methods for describing the trends of these parameters and offering reasonable extrapolations for isotopes with limited experimental data~\cite{von2005systematics,koning2008global}. %Machine learning can be instrumental in addressing this challenge.

In recent years, machine learning has been increasingly applied across the area of nuclear physics, covering various topics such as nuclear theory, experimental methods, accelerator technology, and nuclear data. These results demonstrate its capabilities in data processing and predictions of various measurements. In the realm of nuclear data, machine learning techniques have been widely used as research tools in areas including nuclear mass, nuclear charge radius, separation 
energy, branching ratio of radioactive decay, reaction cross-sections, {\it etc.}~\cite{britton2020ml,boehnlein2022colloquium,he2023machine}. In the study of nuclear level density, Bayesian neural network has been employed to train the level density parameter $a$ from RIPL-2 for CTM, BFM and GSM~\cite{ozdougan2021estimations}. Energy shifts were not involved in this work, and no extrapolation of the data was conducted. However, this is a valuable attempt that demonstrates the feasibility of utilizing machine learning to investigate the variation patterns of level density parameters. On the other hand, Ref.~\cite{von2005systematics} conducted a systematic study and proposed a simple expression related to shell effects to calculate energy shifts. %This expression could, to some extent, describe these variations in parameters. 
It suggests that the changes in the parameter $\varDelta$ follow a pattern that can be captured by machine learning. Among various machine learning models, Feedforward neural network (FNN) is widely utilized by nuclear physicists due to its simple structure, strong interpretability, and powerful data fitting capabilities~\cite{wu2020calculation,yang2021taming,shang2022prediction,yang2023calibration,PhysRevC.108.034311,li2023comparative}. Therefore, we employ the FNN to learn and extrapolate the level density parameter and energy shift of the BFM, to enable the prediction of nuclear level density.

The article is arranged as follows. The details of BFM and FNN are discussed in Sec.~\ref{sec:2}. The Sec.~\ref{sec:3} evaluates the parameters $a$ and $\varDelta$ obtained from the FNN, and employs these parameters for extensive verification calculations, including neutron resonance spacings, cumulative number of levels, and level densities. The Sec.~\ref{sec:4} provides a summary and offers some perspectives.

\section{\label{sec:2}The back-shifted fermi gas model and the feedforward neural network}

\subsection{The back-shifted Fermi gas model}

According to BFM~\cite{dilg1973level,koning2008global}, the total nuclear level density for a given excitation energy $E_{x}$ can be expressed as
\begin{equation}
\rho_{\mathrm{BFM}}^{\mathrm{tot}}(E_{x}) =\left[\frac{1}{\rho(E_{x})}+\frac{1}{\rho_{0}(E_{x})}\right]^{-1},
\end{equation}
where $\rho(E_{x})$ is given by
\begin{equation}
\rho(E_{x})=\frac{1}{\sqrt{2\pi}\sigma}\frac{\sqrt{\pi}}{12}\frac{\mathrm{exp\left[2\sqrt{\mathit{a}\mathit{U}}\right]}}{a^{\frac{1}{4}}U^{\frac{5}{4}}},
\end{equation}
and the expression for the second term is
\begin{equation}
\rho_{0}(E_{x})=\frac{\mathrm{exp}(1)a}{12\sigma}\mathrm{exp}(aU).
\end{equation}
In this context, both the level density parameter $a$ and the spin cut-off parameter $\sigma^2$ are functions of energy, and  the effective excitation energy $U$ is given by $U=E_{x}-\varDelta$.

For the level density parameter $a$, it can be expressed in the following form~\cite{ignatyuk1975phenomenological}
\begin{equation}
a(E_{x})=\tilde{a}\left(1+\delta{W}\frac{1-\mathrm{exp}[-\gamma{U}]}{U}\right).   \label{eq:aE}
\end{equation}
Here, $\tilde{a}$ is the asymptotic level density value that one would obtain in the absence of any shell effects, and it can be parameterized as a function of the mass number $A$, written as
\begin{equation}
\tilde{a}=\alpha A+\beta A^{\frac{2}{3}}.
\end{equation}
$\delta{W}$ is the shell correction term, defined as the difference between the experimental nuclear mass and the liquid-drop model mass. $\gamma$ is a parameter used to describe the decay of shell effects with increasing energy, expressed as
\begin{equation}
    \gamma = \frac{\gamma_{1}}{A^{\frac{1}{3}}}.
\end{equation}
For the energy shift $\varDelta$, the expression is as follows
\begin{equation}
    \varDelta =\begin{cases}
    \frac{12}{\sqrt{A}}+\delta,  &\text{for\quad even-even}, \\
   -\frac{12}{\sqrt{A}}+\delta,  &\text{for\quad odd-odd}, \\
    \delta,                      &\text{for\quad odd-A}. \\
\end{cases}\label{eq:delta}
\end{equation}

In the calculations of this study, for convenience, the spin-cutoff parameter $\sigma^2$ is expressed in the form of the following piecewise function
\begin{equation}
\sigma^2=\begin{cases}
\sigma_{d}^{2}, &\text{for}\quad E_{x} \leq \varDelta,\\
\sigma_{d}^{2}+\frac{E_{x}-\varDelta}{S_{n}-\varDelta}(\sigma_{F}^{2}(E_{x})-\sigma_{d}^{2}),&\text{for}\quad \varDelta < E_{x} < S_{n},\\
\sigma_{F}^{2}(E_{x}),&\text{for}\quad  E_{x} \geq S_{n},\\
\end{cases}
\end{equation}
where $\sigma_{d}^{2}$ is represented as
\begin{equation}
\sigma_{d}^{2}=\left(0.83A^{0.26}\right)^{2},
\end{equation}
and $\sigma_{F}^{2}(E_{x})$ is 
\begin{equation}
\sigma_{F}^{2}(E_{x})=0.01389\frac{A^{5/3}}{\tilde{a}}\sqrt{\mathit{a}\mathit{U}}.
\end{equation}

Finally, taking into account the parity and spin distributions, the level density is expressed as
\begin{equation}
\rho{(U,J,P)}=\frac{1}{2}\frac{2J+1}{2\sigma^2}\mathrm{exp}\left[-\frac{(J+\frac{1}{2})^2}{2\sigma^2}\right]\rho^{\mathrm{tot}}_{\mathrm{BFM}}(E_{x}).
\end{equation}
On this basis, neutron resonance spacings and the cumulative number of levels can also be calculated.

\subsection{The feedforward neural network}
The FNN consist of multiple layers, including the input layer, hidden layers, and output layer. Each layer contains multiple neurons. Neurons receive incoming information and pass it to the next layer through an activation function. The information ultimately reaches the output layer, completing the data processing.

In this study, the hyperbolic tangent function (tanh) is chosen as the activation function for the FNN hidden layer, and the mean square error (MSE) is selected as the loss function for the network, expressed as
\begin{equation}
    Loss=\frac{1}{N_{t}}\sum^{N_{t}}_{i=1}(y_{tar}-y_{pre})^{2},
\end{equation}
where $N_{t}$ represents the size of the training set, $y_{tar}$ represents the fitted values, and $y_{pre}$ represents the predictions given by the FNN. During the network training process, the stochastic gradient descent method~\cite{robbins1951stochastic} is applied to continuously update the network parameters in order to minimize the loss function.

The data utilized in this study comprises level density parameters at the neutron separation energy $a(S_{n})$ and energy shifts $\varDelta$ for 289 nuclei. These data are obtained by fitting the $D_{0}$ and discrete levels in RIPL-2~\cite{belgya2006handbook,koning2008global}. The corresponding dataset can be found in the data files of TALYS version 1.8 and above~\cite{koning2012modern}. It is important to note that $\delta$ values are provided for an additional 846 nuclei with only discrete level data. However, to ensure the consistency of the utilized data, these values were not used to train the neural network.

In view of the intricate relationship between $a(S_{n})$ and $\delta$, and the overall lack of strong regularity in $\delta$, the dataset underwent classification. Utilizing Eqs.~(\ref{eq:delta}) to transform $\delta$ back to $\varDelta$, the training of the network was conducted separately for $a(S_{n})$ and $\varDelta$. Additionally, the $\varDelta$ dataset was categorized into three classes: even-even nuclei, odd-odd nuclei, and odd-$A$ nuclei, and each was used for training the network. For $a(S_{n})$, it explicitly includes the neutron separation energy $S_{n}$ in Eq.~(\ref{eq:aE}), and when expressed as a function of the mass number $A$, it exhibits pronounced shell effects. Therefore, the proton number $Z$, neutron number $N$, the difference between each of them and the nearest magic numbers ($V_{z}$, $V_{n}$), as well as the neutron separation energy $S_{n}$, are chosen as input variables for the FNN. As for $\varDelta$, the selected input variables for the network are $Z$, $N$, $V_{z}$, and $V_{n}$.

During training, the data is randomly divided into a training set and a validation set, with the training set accounting for $80\%$ of the data and the validation set for $20\%$. FNN parameters are randomly initialized for each training iteration. In order to capture all features of the original data while minimizing the MSE, each network undergoes 1000 repetitions of training. The final results for $a(S_{n})$ and $\varDelta$ are obtained by averaging the 100 sets of data provided by the FNN, with the smallest MSE.

\section{\label{sec:3}Results and discussion}

\begin{figure}[b]
\centering
\includegraphics[width=8cm]{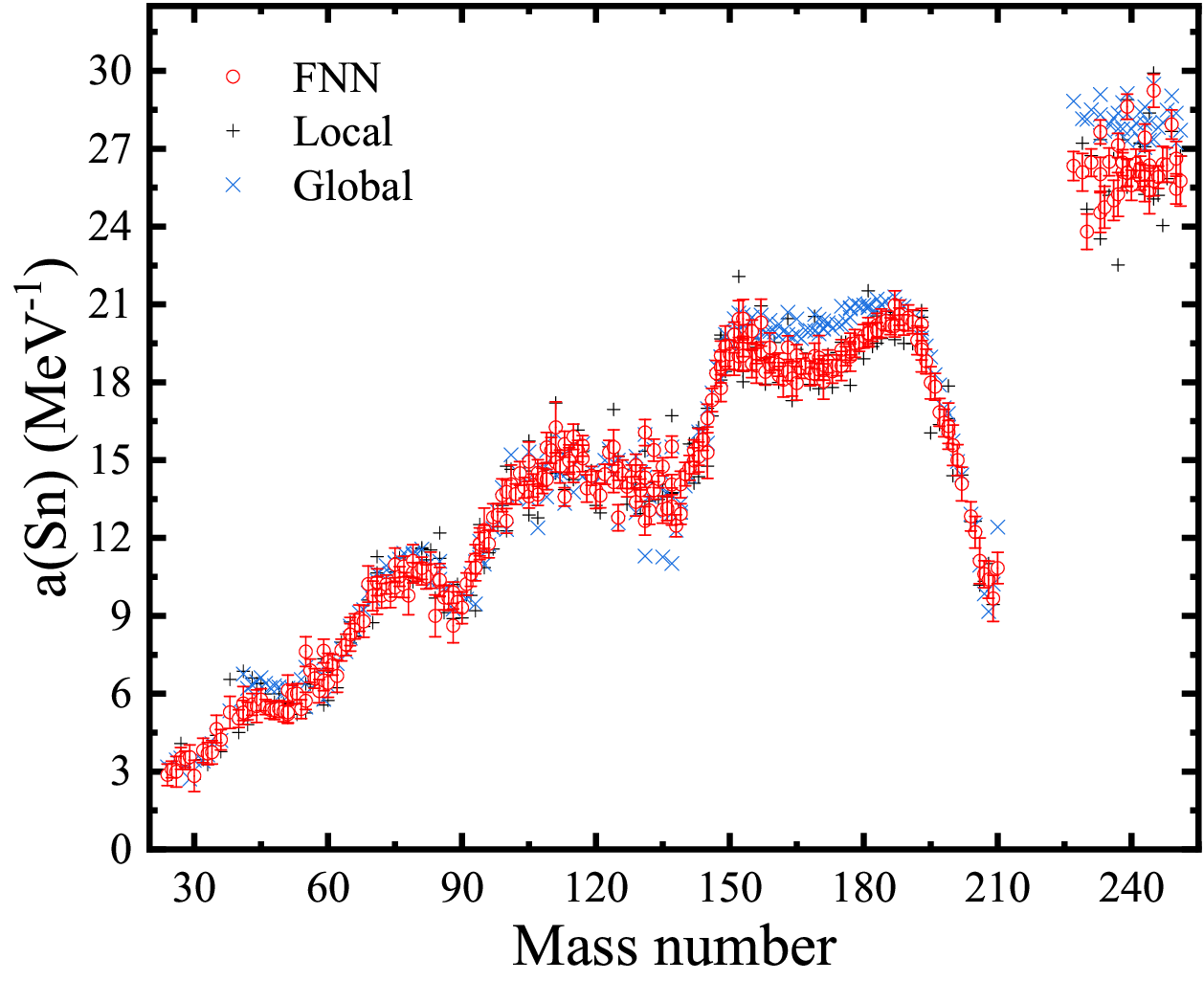}% Here is how to import EPS art
\caption{\label{fig:epsart}(Color online) Level density parameters $a$ at the neutron separation energy $S_{n}$. The parameters obtained from the FNN (red circles with error bars) and global parameters (blue crosses) were compared with the local parameters (black crosses).}\label{fig:asn}
\end{figure}

Figure~\ref{fig:asn} depicts the values of $a(S_{n})$ provided by the FNN for all 289 nuclei. In addition, the results of parameters fitted to experimental data (local parameters) and global parameters are also presented for comparison. The error bars in the FNN results are derived from twice the standard deviation of multiple training outcomes, indicating that approximately $95\%$ of the data falls within the range of the error bars. It is evident that the FNN effectively captures the variation patterns in the local parameters $a(S_{n})$, especially for nuclei with mass number above 150, where global parameters clearly provide overestimated predictions. Overall, the FNN yields a root mean square deviation of 0.67, significantly outperforming the global parameters with a deviation of 1.18.

\begin{figure*}
\centering
\includegraphics[width=16cm]{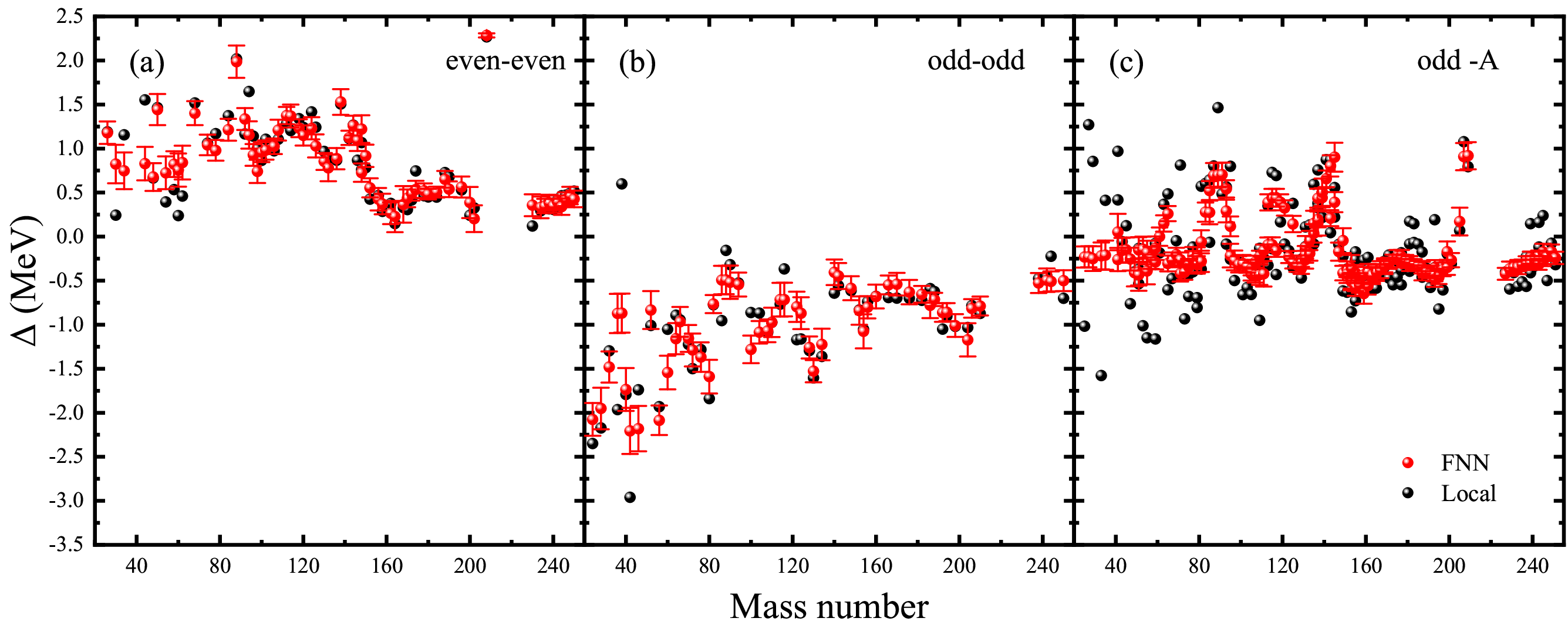}% Here is how to import EPS art
\caption{\label{fig:wide}The energy shift $\varDelta$ for even-even nuclei, odd-odd nuclei, and odd-$A$ nuclei. The parameters obtained from the FNN (red spheres with error bars) were compared with the local values (black spheres).}\label{fig:delta}
\end{figure*}

\begin{figure*}
\centering
\includegraphics[width=16cm]{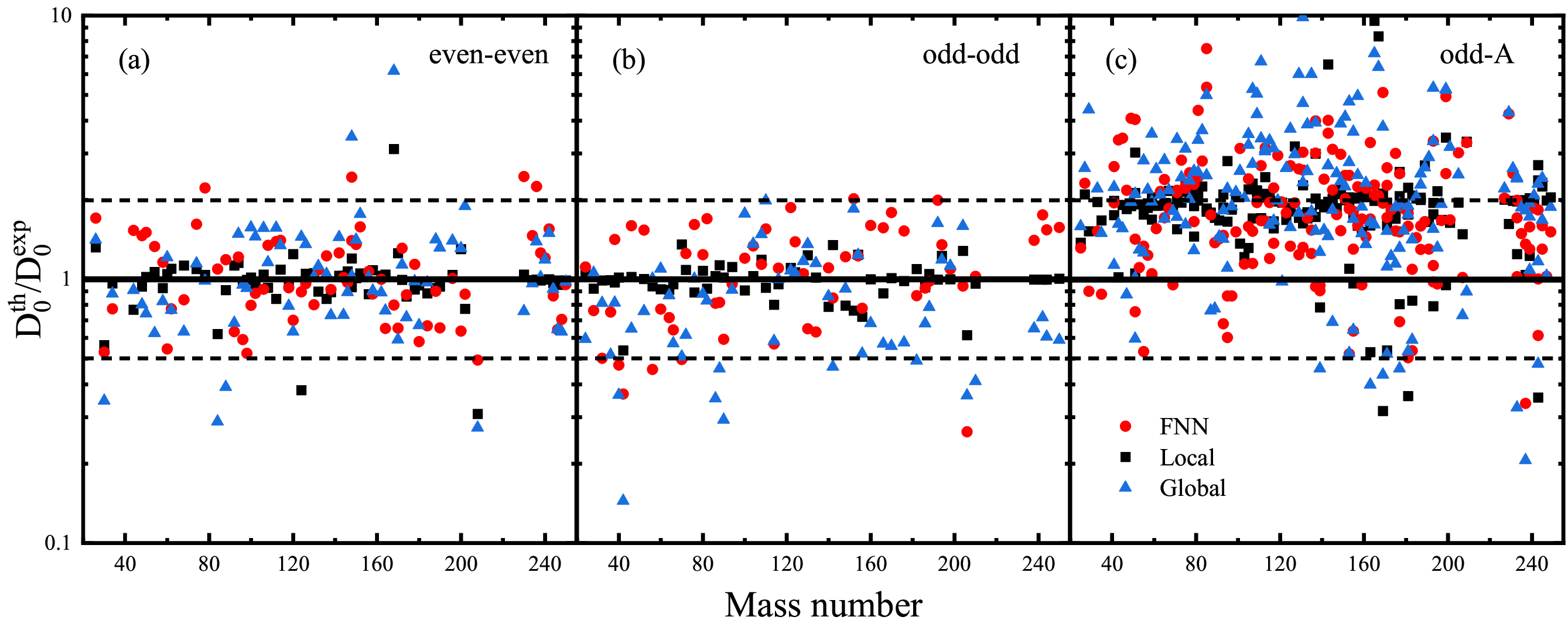}% Here is how to import EPS art
\caption{\label{fig:wide}(Color online) The ratio of the $s$-wave neutron resonance spacings $D_{0}$ predicted by the BFM for even-even, odd-odd, and odd-$A$ nuclei to the experimental data from RIPL-2 as a function of mass number $A$. The parameters obtained from the FNN (red circles) and global parameters (blue triangles) were compared with the local parameters (black squares).}\label{fig:d0}
\end{figure*}

\begin{table}[b]%The best place to locate the table environment is directly after its first reference in text
\caption{\label{tab:table1}%
The root mean square deviation factor $f_{\textrm{rms}}$ for $D_{0}$ values obtained through BFM calculations using FNN parameters, local parameters and global parameters, in comparison to the experimental data in RIPL-2.
}
\begin{ruledtabular}
\begin{tabular}{ccccc}
\textrm{Parameter}&
$f_{\textrm{rms}}^{\textrm{even-even}}$&
$f_{\textrm{rms}}^{\textrm{odd-odd}}$&
$f_{\textrm{rms}}^{\textrm{odd-$A$}}$&
$f_{\textrm{rms}}^{\textrm{Entirety}}$\\
\textrm{types}&
\textrm{67 nuclei}&
\textrm{56 nuclei}&
\textrm{165 nuclei}&
\textrm{288 nuclei}\\
\colrule
FNN & 1.46 & 1.58 & 2.04 & 1.87\\
Local & 1.31 & 1.18 & 2.05 & 1.76\\
Global & 1.62 & 1.76 & 2.59 & 2.22\\
\end{tabular}
\end{ruledtabular}
\end{table}

\begin{figure}
\centering
\includegraphics[width=8cm]{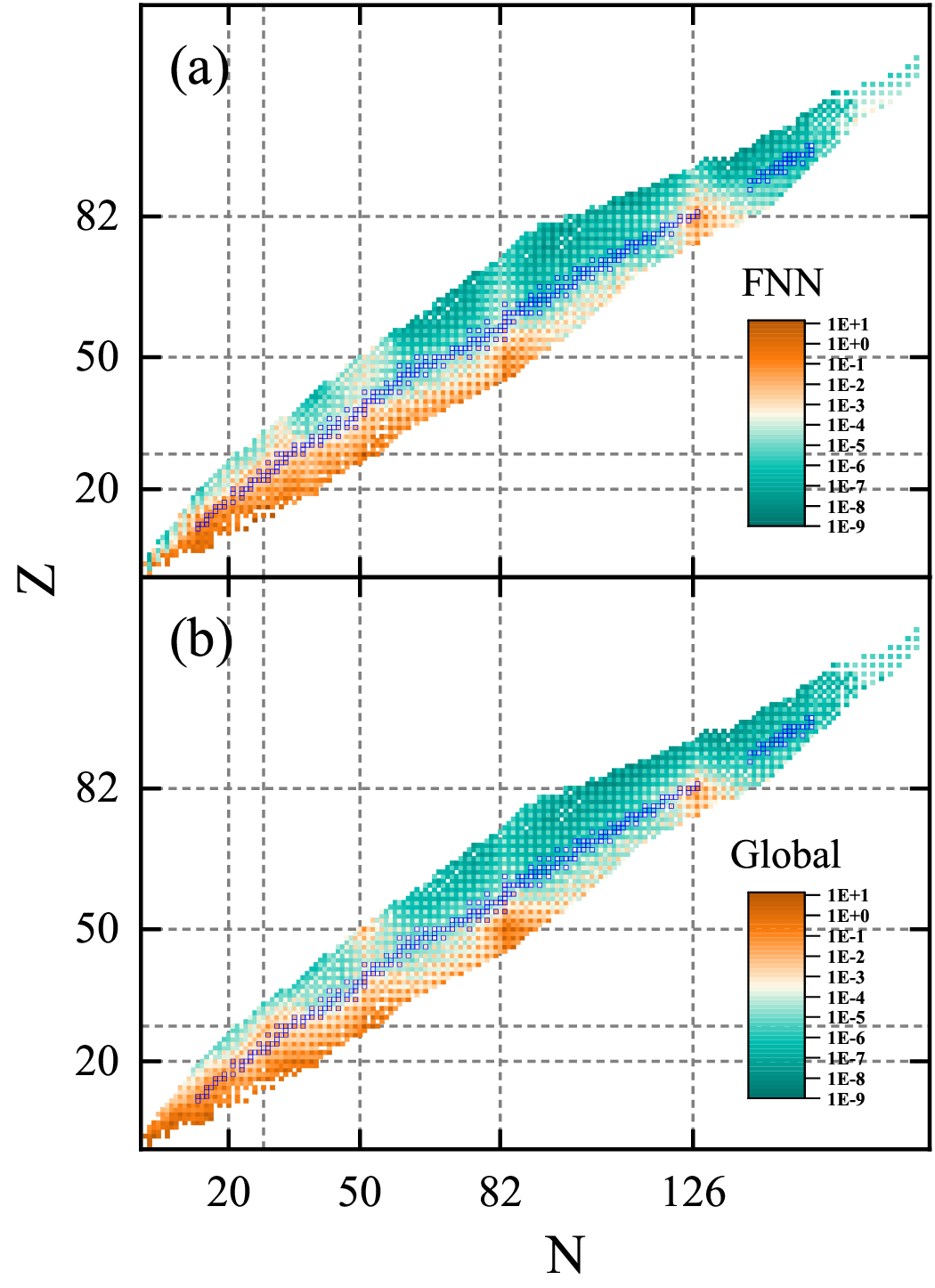}% Here is how to import EPS art
\caption{\label{fig:wide}(Color online) The $D_{0}$ values for 3171 nuclei, calculated using the parameters obtained from the FNN and global parameters, including 289 nuclei within the training set (represented by blue boxes) and 2882 nuclei outside the training set.}\label{fig:waitui}
\end{figure}
\begin{figure}
\centering
\includegraphics[width=8cm]{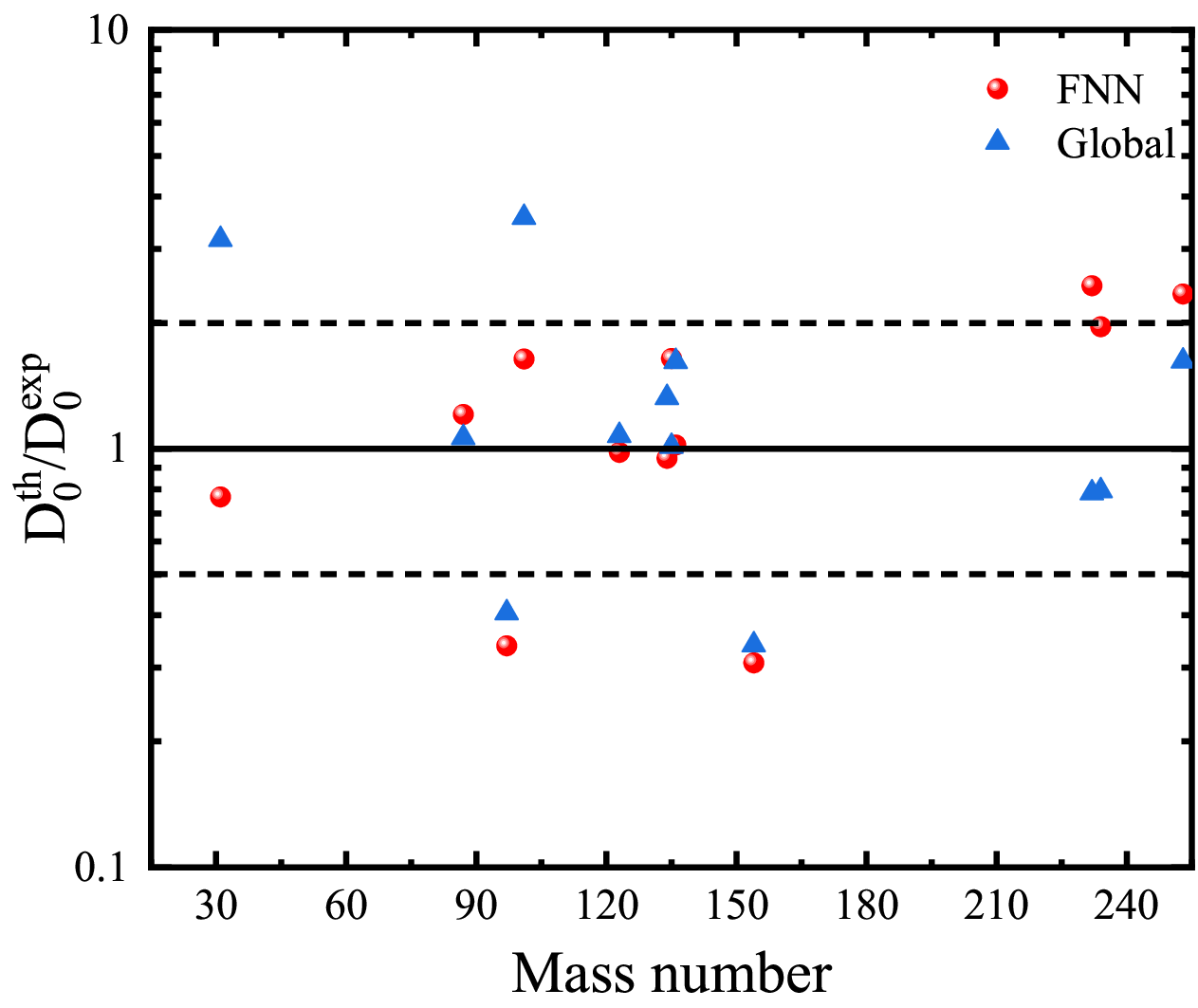}% Here is how to import EPS art
\caption{\label{fig:wide}(Color online) The ratio of the $D_{0}$ values predicted by BFM for 12 nuclei outside the training set to the experimental data in RIPL-3. The results obtained from FNN parameters (red spheres) were compared with the results from global parameters (blue triangles).}\label{fig:check}
\end{figure}

\begin{figure*}
\centering
\includegraphics[width=16cm]{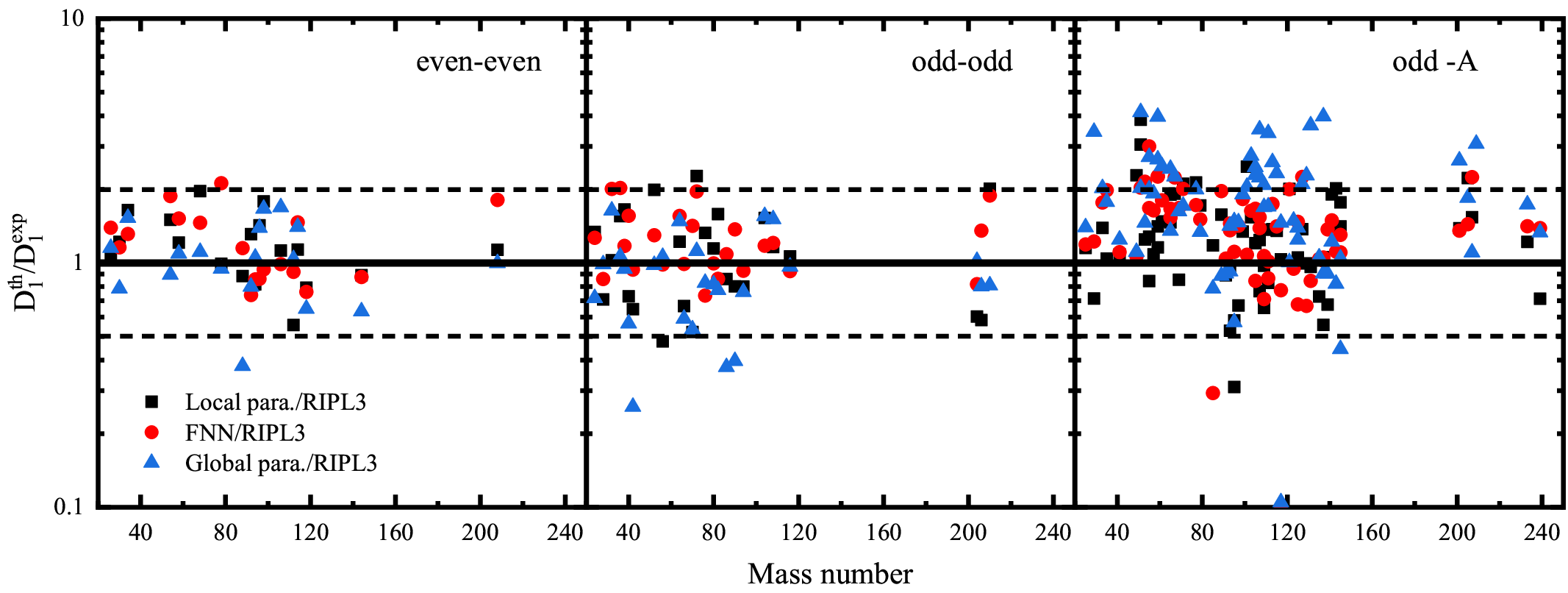}% Here is how to import EPS art
\caption{\label{fig:wide}(Color online) The ratio of the $p$-wave neutron resonance spacings $D_{1}$ predicted by the BFM for even-even, odd-odd, and odd-$A$ nuclei to the experimental data from RIPL-3 as a function of mass number $A$. The FNN parameters (red circles) and global parameters (blue triangles) were compared with the local parameters (black squares).}\label{fig:d1}
\end{figure*}

\begin{table}[b]%The best place to locate the table environment is directly after its first reference in text
\caption{\label{tab:table2}%
The root mean square deviation factor $f_{\textrm{rms}}$ for $D_{1}$ values calculated using different types of parameters compared to the experimental data in RIPL-3.
}
\begin{ruledtabular}
\begin{tabular}{ccccc}

\textrm{Parameter}&
$f_{\textrm{rms}}^{\textrm{even-even}}$&
$f_{\textrm{rms}}^{\textrm{odd-odd}}$&
$f_{\textrm{rms}}^{\textrm{odd-$A$}}$&
$f_{\textrm{rms}}^{\textrm{Entirety}}$\\
\textrm{types}&
\textrm{18 nuclei}&
\textrm{26 nuclei}&
\textrm{72 nuclei}&
\textrm{116 nuclei}\\
\colrule
FNN & 1.42 & 1.58 & 1.90 & 1.76\\
Local & 1.40 & 1.41 & 2.01 & 1.79\\
Global & 1.45 & 1.61 & 2.13 & 1.92\\
\end{tabular}
\end{ruledtabular}
\end{table}

\begin{figure}
\centering
\includegraphics[width=8cm]{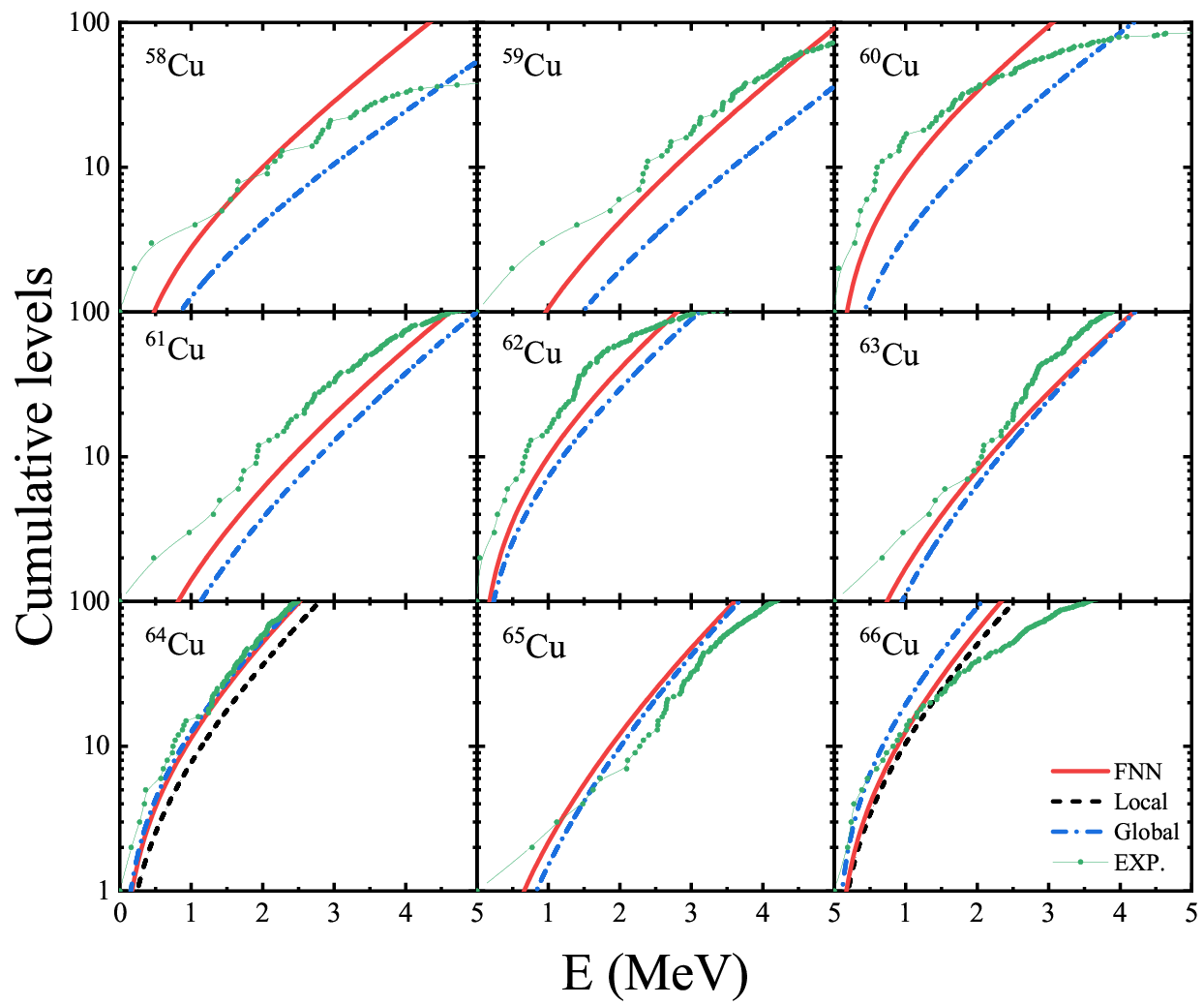}% Here is how to import EPS art
\caption{\label{fig:wide}(Color online) Cumulative number of levels for $^{58-66}\textrm{Cu}$. Comparison was made between the results calculated from FNN parameters (red line), global parameters (blue dotted dash line), and local parameters (black dush line), and the experimental data (green dot line).}\label{fig:cu}
\end{figure}
\begin{figure}
\centering
\includegraphics[width=8cm]{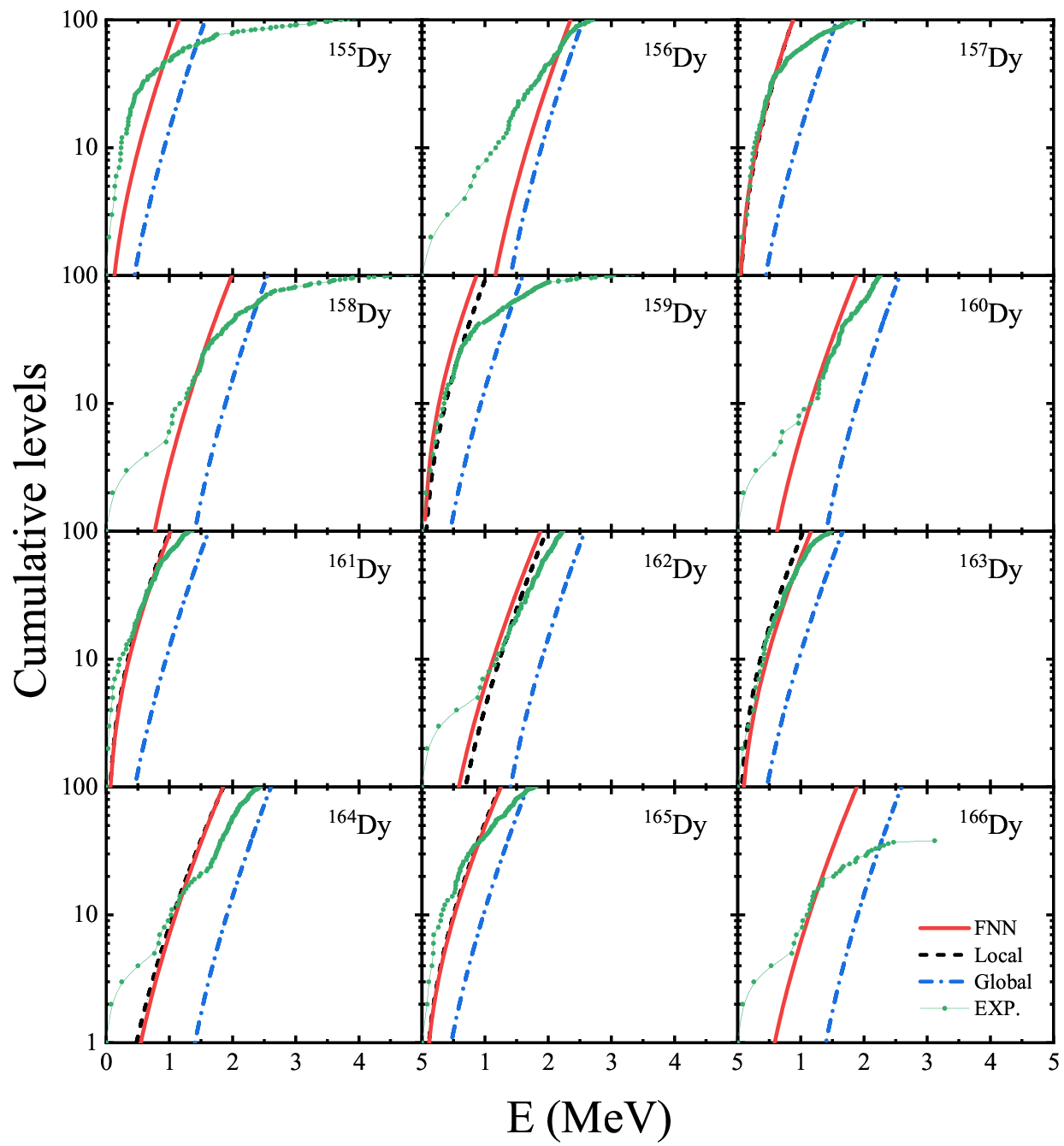}% Here is how to import EPS art
\caption{\label{fig:wide}(Color online) The cumulative number of levels calculated with FNN parameters, local parameters, and global parameters are compared with the experimental data, for $^{155-166}\textrm{Dy}$.}\label{fig:dy}
\end{figure}
\begin{figure}
\centering
\includegraphics[width=8cm]{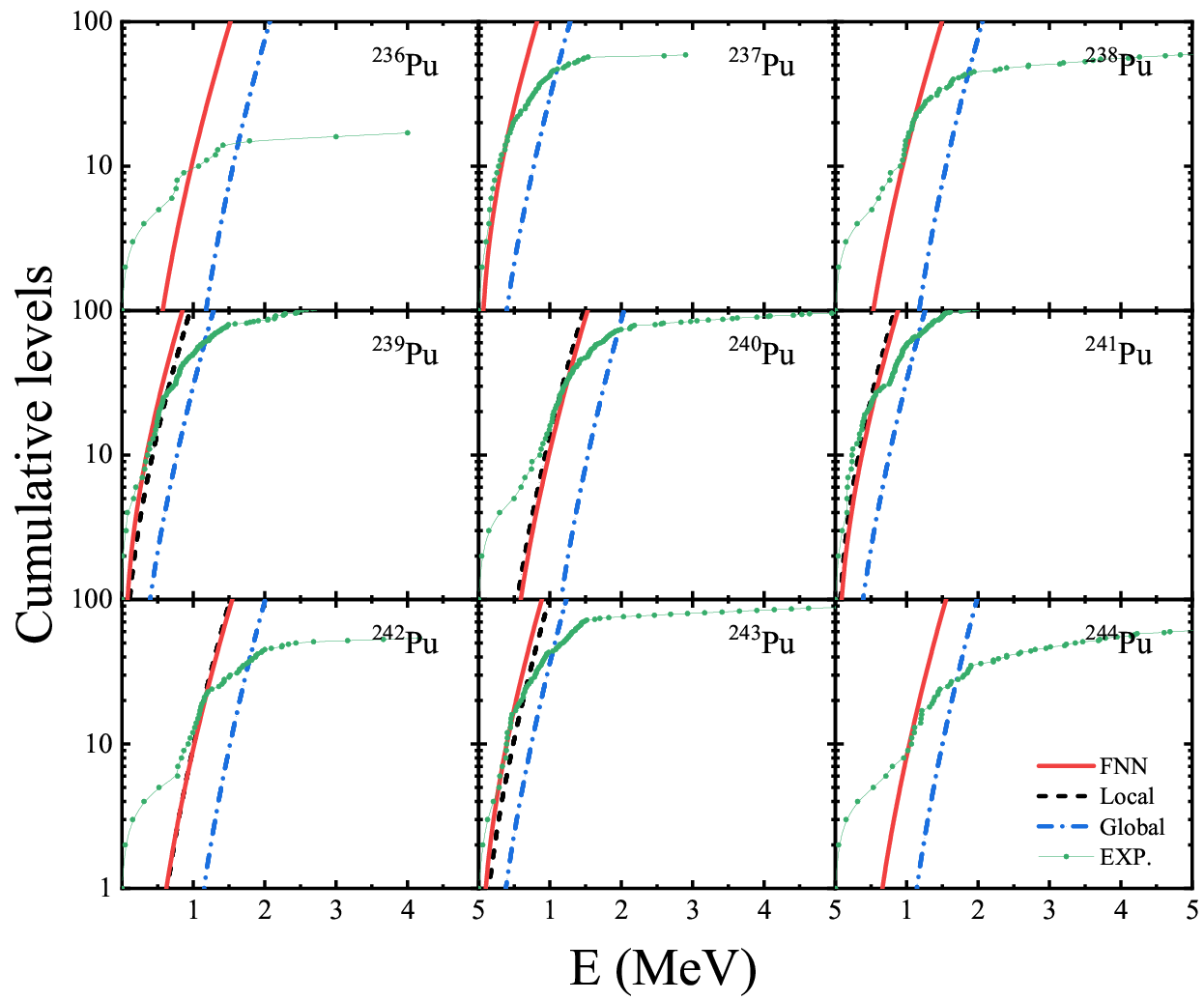}% Here is how to import EPS art
\caption{\label{fig:wide}(Color online) The cumulative number of levels calculated with FNN parameters, local parameters, and global parameters are compared with the experimental data, for $^{236-244}\textrm{Pu}$.}\label{fig:pu}
\end{figure}
\begin{figure}
\centering
\includegraphics[width=8cm]{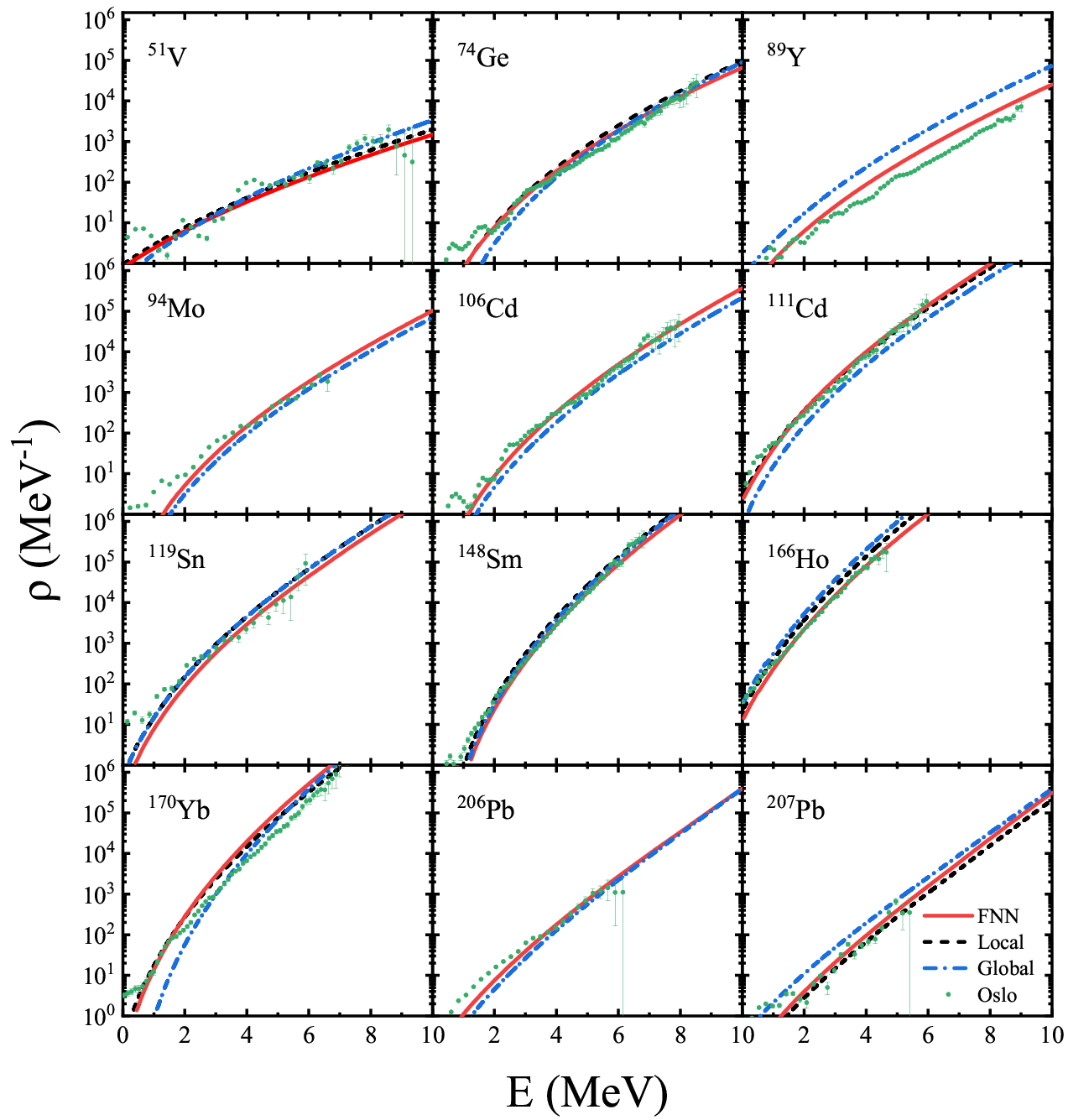}% Here is how to import EPS art
\caption{\label{fig:wide}(Color online) Comparison among the total NLDs calculated using FNN parameters (red line), local parameters (black dash line), and global parameters (blue dotted dush line) in the BFM with experimental data (green dots with error bars).}\label{fig:nld}
\end{figure}
Regarding $\varDelta$, the results of FNN training are presented in Fig.~\ref{fig:delta}. Error bars are also provided at two times the standard deviation. In the dataset, there are 67 even-even nuclei, 56 odd-odd nuclei, and 166 odd-$A$ nuclei. The FNN yields root mean square deviations of 0.20, 0.33, and 0.36 for above categories, respectively. It can be observed that, whether it is an even-even, odd-odd or odd-$A$ nuclei, the absolute value of $\varDelta$ continuously oscillates and tends towards zero with an increase in the mass number. Meanwhile, variations in the values of $\varDelta$ are evident around magic numbers. This underscores a clear fact that the Fermi gas model is better suited for describing systems with a larger number of nucleons. For light nuclei, however, larger parameter adjustments are necessary to align with experimental data. In general, for both even-even and odd-odd nuclei, the FNN effectively reproduces the changing trends of $\varDelta$. The inclusion of magic number information in the network inputs allows FNN to capture the associated variations. For instance, in Fig.~\ref{fig:delta} (a), the case of $^{\mathrm{208}}\mathrm{Pb}$ clearly deviates from the overall trend, but FNN accurately describes its results. Another notable fact is that FNN's description of $\varDelta$ for medium-heavy nuclei is significantly better than for light nuclei. The primary reason is the extremely pronounced variations of $\varDelta$ in the light nuclei region, and these variations do not exhibit a clear connection to magic number information. Additionally, the amount of data used to train the network in the light nuclei region is relatively limited. For example, in Fig.~\ref{fig:delta} (b), the case of $^{\mathrm{38}}\mathrm{Cl}$ exhibits local values that are positive, making it challenging for FNN to capture such variations. Especially for odd-$A$ nuclei, FNN can roughly describe the patterns of local values in the region above mass number 80, but struggles to capture these oscillations below mass number 80.

$D_{0}$ is the most reliable experimental data related to NLD. Figure~\ref{fig:d0} depicts the ratio of calculated $D_{0}$, using parameters obtained from FNN, to the corresponding experimental values from RIPL-2. Meanwhile, the results of calculations using local parameters and global parameters are also given separately. Due to the absence of $D_{0}$ for $^{251}\mathrm{Cf}$ in RIPL-3, subsequent comparisons have not taken this nucleus into consideration. It can be observed that, for both even-even and odd-odd nuclei, the calculated results for all three sets of parameters mostly fall within a range of 0.5 to 2 times the experimental values. For odd-$A$ nuclei, the calculated results using local parameters mostly cluster around 2 times the experimental values, while the results from FNN parameters and global parameters are scattered above and below the dashed line representing 2 times the experimental values. 

The differences in the calculated results for the three types of parameters, i.e., local, global and FNN are revealed through the root mean square deviation factor, denoted as $f_{\textrm{rms}}$, between the theoretical and experimental values of $D_{0}$. The $f_{\textrm{rms}}$ is defined as follows~\cite{hilaire2001combinatorial}:
\begin{equation}
f_{\textrm{rms}}=\mathrm{exp}{\left[\frac{1}{N_e}\sum^{N_e}_{i=1}{ln^2\frac{D^{th}_{0}}{D^{exp}_{0}}}\right]}^{1/2}.\label{eq:rms}
\end{equation}
The corresponding results are provided in Table~\ref{tab:table1}. In an overall assessment, the parameters obtained from FNN yield a value of $f_{\textrm{rms}}=1.87$, which is bigger than the local parameters (1.76) but superior to the global parameters (2.22). This result is consistent with its performance in both even-even and odd-odd nuclei. However, for odd-$A$ nuclei, the parameters from FNN yield a value of $f_{\textrm{rms}}=2.04$, even slightly surpassing the local parameters (2.05) and significantly outperforming the global parameters (2.59). From Fig.~\ref{fig:d0} (c), it can be observed that the results from FNN are more scattered, with a larger proportion falling below the dashed line representing twice the experimental values.

Utilizing FNN, a comprehensive extrapolation was performed for $a(S_{n})$ and $\varDelta$ over a wide range, providing parameters for 2882 nuclei beyond the training set. Specific parameters are provided in the Supplemental Material~\cite{supplementalmaterial}, with the extrapolation range covering $Z$ from 2 to 118 and $N$ from 1 to 177. In the input of FNN, $Z=126$ and $N=184$ are set to magic numbers. The ground-state information for all these nuclei can be found in RIPL-3, and this information is assumed to be entirely correct. Subsequently, this ground-state level properties, along with the parameters obtained from FNN, are used to calculate $D_{0}$. The corresponding results are presented in Fig.~\ref{fig:waitui} (a). For comparison, the calculated results using global parameters are provided in Fig.~\ref{fig:waitui} (b). In comparing the results from the two types parameter, both present a similar pattern: within an isotopic chain, $D_{0}$ generally increases with the neutron number, and nuclei near magic numbers exhibit significantly larger $D_{0}$ values compared to other nuclei. Comparing Fig.~\ref{fig:waitui} (a) with Fig.~\ref{fig:waitui} (b), there is a noticeable difference around $Z=50$, $N=50$ ($^{100}\mathrm{Sn}$). Global parameters predict a larger $D_{0}$ in this region. By comparing the values of $a(S_{n})$ obtained through the two methods, in the vicinity of $^{100}\mathrm{Sn}$, the $a(S_{n})$ values calculated using global parameters tend to be smaller than the predictions from FNN. This difference will lead to larger values of $D_{0}$. The reason for this phenomenon could be the stronger shell correction ($\delta$$W$) in this region. These variations beyond the training set data are entirely beyond the grasp of FNN. The true nature is yet to be revealed through subsequent experimental observations.

RIPL-3 has incorporated $D_{0}$ data for more nuclei compared to RIPL-2, and also provided updates to the $D_{0}$ values for some previously included nuclei~\cite{capote2009ripl}. The presence of $D_{0}$ values for 12 nuclei in RIPL-3, not included in the training set, provides an opportunity to evaluate the validity of the parameters from FNN. The target nuclei corresponding to these $D_{0}$ values include $^{30}\mathrm{Si}$, $^{86}\mathrm{Kr}$, $^{96}\mathrm{Zr}$, $^{100}\mathrm{Ru}$, $^{122}\mathrm{Sn}$, $^{134}\mathrm{Cs}$, $^{135}\mathrm{Cs}$, $^{133}\mathrm{Ba}$, $^{153}\mathrm{Gd}$, $^{231}\mathrm{Pa}$, $^{233}\mathrm{Pa}$ and $^{252}\mathrm{Cf}$. The ratios of $D_{0}$ values calculated using FNN to experimental data are presented in Fig.~\ref{fig:check}. The results obtained using global parameters are also given simultaneously. Overall, the results from FNN are distributed in the range of 0.5 to 2, with the minimum value being 0.31 ($^{154}\mathrm{Gd}$) and the maximum value being 2.45 ($^{232}\mathrm{Pa}$). In comparison to global parameters, FNN's results are superior for nuclei with mass numbers below 200, while, FNN's results are slightly inferior for the three nuclei with mass numbers above 200. In general, the parameters obtained from FNN yield a value of $f_{\textrm{rms}}=1.98$, and the global parameters similarly give 1.98 in this context. Despite the $D_{0}$ updates in RIPL-3 compared to RIPL-2, the parameters from FNN consistently yield $f_{\textrm{rms}}=1.98$ for the remaining 288 nuclei provided in RIPL-3. The obtained result is consistent with the outcomes for the predicted 12 nuclei. This, to some extent, indicates that the approach of extrapolating parameters using FNN is stable, at least for nuclei close to the training set. In contrast, the global parameters yield $f_{\textrm{rms}}=2.19$ for the remaining 288 nuclei in RIPL-3.

Moreover, the $p$-wave neutron resonance spacing $D_{1}$ also can be used to evaluate the quality of parameters. RIPL-3 provides $D_{1}$ values for 116 nuclei, including 18 even-even nuclei, 26 odd-odd nuclei, and 72 odd-$A$ nuclei. The ratio of the calculated $D_{1}$ using parameters obtained from FNN to the experimental data from RIPL-3 is presented in Fig.~\ref{fig:d1}. The results from local parameters and global parameters are also displayed for comparison. It can be observed that, unlike the case of $D_{0}$, the results of calculations using local parameters are no longer widely distributed along lines representing 1 or 2. Instead, similar to FNN parameters and global parameters, they are scattered between 0.5 and 2 times the experimental values. Similar to Eq.~(\ref{eq:rms}), the corresponding root mean square deviation is provided in Table~\ref{tab:table2}. For even-even and odd-odd nuclei, the calculated results from FNN parameters yield $f_{\textrm{rms}}$ bigger than local parameters but smaller than global parameters. However, for odd-$A$ nuclei, the performance of FNN parameters is better than that of local parameters. In the end, for all 116 nuclei, FNN parameters yield $f_{\textrm{rms}}=1.76$, which is slightly better than the 1.79 from local parameters and superior to the 1.92 from global parameters.

Discrete levels at low excitation energies are the most abundant experimental data related to NLD. In the process of fitting BFM parameters, it is observed that the cumulative number of levels is highly sensitive to variations in energy shift. Certainly, the cumulative count of discrete levels can be employed to validate the reasonableness of $\varDelta$ obtained from FNN. Three isotopic chains, including Cu, Dy and Pu, have been selected to validate the parameters obtained using FNN. 
Figure~\ref{fig:cu} displays the results of the cumulative number of levels from FNN for nuclei ranging from $^{58}\mathrm{Cu}$ to $^{66}\mathrm{Cu}$. The corresponding results of global parameter are also provided. Additionally, for nuclei within the training set, results obtained using local parameters are presented. For the Cu isotopic chain, only $^{64}\mathrm{Cu}$ and $^{66}\mathrm{Cu}$ are within the training set. From Fig.~\ref{fig:cu}, it can be observed that the parameters provided by FNN fit well with the experimental data for $^{64}\mathrm{Cu}$ and $^{66}\mathrm{Cu}$. For $^{64}\mathrm{Cu}$, the results even surpass the local parameters. Besides, for nuclei outside the training set, FNN yields results superior to global parameters, particularly for $^{58-60}\mathrm{Cu}$, which are more distant from the training set, demonstrating a good description by FNN. 

Similar to Fig.~\ref{fig:cu}, Fig.~\ref{fig:dy} displays the calculated results of the cumulative number of levels for $^{155}\mathrm{Dy}$ to $^{166}\mathrm{Dy}$. For nuclei within the training set ($^{157}\mathrm{Dy}$, $^{159}\mathrm{Dy}$, $^{161-165}\mathrm{Dy}$), the results from FNN parameters are very close to those obtained with local parameters. Regarding the extrapolation results, FNN demonstrates strong predictive capabilities, effectively reproducing the experimental results for $^{158}\mathrm{Dy}$, $^{160}\mathrm{Dy}$ and $^{166}\mathrm{Dy}$. Moreover, for $^{155}\mathrm{Dy}$ and $^{156}\mathrm{Dy}$, FNN provides results superior to global parameters. The results for Pu isotopes from $^{236}\mathrm{Pu}$ to $^{244}\mathrm{Pu}$ are presented in Fig.~\ref{fig:pu}. It can be observed that, for nuclei within the training set ($^{239-243}\mathrm{Pu}$), the parameters from FNN provide results that closely overlap with those from local parameters while effectively reproducing experimental data. Additionally, for extrapolated nuclei ($^{236-238}\mathrm{Pu}$, $^{244}\mathrm{Pu}$), FNN parameters yield excellent agreement with experimental data. Overall, the parameters from FNN provide a good description of the cumulative number of levels for nuclei in these three chains, especially for nuclei outside the training set where FNN still provides accurate predictions. This serves as strong evidence for the reliability of the $\varDelta$ obtained from FNN.

Finally, the comparison between the calculated results of the NLD using FNN parameters for the 12 nuclei between $^{51}\mathrm{V}$ to $^{207}\mathrm{Pb}$ and Oslo data~\cite{larsen2006microcanonical,renstrom2016low,larsen2016experimentally,utsunomiya2013photoneutron,larsen2013transitional,toft2010level,siem2002level,pogliano2023observation,agvaanluvsan2004level,syed2009level} is presented in Fig.~\ref{fig:nld}. The results obtained using global parameters are also shown, and the results from local parameters are retained for nuclei within the training set. It can be observed that, within the framework of the BFM, parameters from FNN effectively describe the NLD for nuclei both within and beyond the training set. However, due to the limitations of the Fermi gas model, the parameters from the FNN were unable to capture the details of the NLD variations with excitation energy observed in the Oslo data. For example, in Fig.~\ref{fig:nld}, the experimental NLD data for $^{51}\mathrm{V}$ exhibit oscillations with increasing excitation energy, a pattern that is challenging for the BFM to describe.

\section{\label{sec:4}SUMMARY AND PROSPECTS}

The FNN has been employed to predict parameters of the back-shifted Fermi gas model, including the level density parameter at neutron separation energy $a(S_{n})$ and the energy shift $\varDelta$. Considering the convenience of data extrapolation, the input for the FNN includes only the proton number $Z$, neutron number $N$, and the corresponding difference to the nearest magic number (shell information). For $a(S_{n})$, the input is further augmented with the neutron separation energy $S_{n}$. The results demonstrate that the FNN effectively reproduces the variation trends of $a(S_{n})$. For the energy shift $\varDelta$, due to significant differences observed among even-even nuclei, odd-odd nuclei, and odd-$A$ nuclei, three separate neural networks are employed to individually train on each category. The results reveal that the FNN effectively captures the patterns of $\varDelta$ variation from intermediate to heavy nuclei. However, for lighter nuclei, the data on the training set is quite limited, concurrently exhibiting irregular variations, which results in FNN providing less precise descriptions of $\varDelta$. This is particularly evident for odd-$A$ nuclei, where the $\varDelta$ values for nuclei with mass numbers below 80 appear to lack a discernible pattern. This suggests that, at least for lighter nuclei, the FNN may benefit from incorporating additional physical information in the input variables to apply soft constraints on its predictions for $\varDelta$. Thanks to the simplicity of the FNN input, extrapolating data becomes straightforward. The FNN provides $a(S_{n})$ and $\varDelta$ values for nearly 3000 nuclei (the detailed parameters can be found in the Supplemental Material~\cite{supplementalmaterial}). Certainly, for nuclei that are significantly distant from the training set, caution must be taken when using these parameters.

The calculations of $D_{0}$ and $D_{1}$ using the parameters obtained from FNN indicate that this set of parameters can effectively describe the experimental data. For $D_{0}$ and $D_{1}$, FNN obtains values of $f_{\textrm{rms}}$ as 1.87 and 1.76, respectively. The results suggest that the parameters from FNN achieve a performance close to that of local parameters. The parameters from FNN also provide a satisfactory description for the 12 nuclei beyond the training set in RIPL-3, yielding a result of $f_{\textrm{rms}}=1.98$. This suggests that the method of providing extrapolated parameters is, at the very least, stable for nuclei close to the training set. For extrapolated nuclei, in a broader comparison, FNN's $D_{0}$ calculation results exhibit patterns similar to those obtained with global parameters. However, there are noticeable differences in certain regions, such as lighter proton-rich nuclei and around $^{100}\mathrm{Sn}$, where FNN consistently yields smaller results. The calculated results for the cumulative number of levels and NLDs also provide a reasonable description of the experimental data, further affirming the reliability of these parameters.

Overall, in this study, FNN has been employed to discover four sets of function relationships with indeterminate specific forms, enabling a proficient description of existing experimental data through the utilization of exceedingly simple physical information. In the future, improvements could be achieved by incorporating additional physical information as inputs to the neural network or by considering the utilization of experimental data to impose constraints on the network, aiming for a more accurate description of the parameters.

\begin{acknowledgments}
This work was supported by the Natural Science Foundation of Jilin Province (No.20220101017JC), National Natural Science Foundation of China (No. 11675063), and Key Laboratory of Nuclear Data Foundation (JCKY2020201C157).
\end{acknowledgments}

\nocite{*}

\bibliography{apssamp}% Produces the bibliography via BibTeX.

\end{document}